# Constraining the Superconducting Gap of Monolayer $FeSe/SrTiO_3$ via Statistical Analysis and Tip-Height-Dependent Tunneling Spectroscopy

*Dongwon Shin[1,2], Xiyou Peng[1], Kaleb McClure[1], Paolo Vilmercati[1], Chris Howard[1,2], Petro Maksymovych[3], Norman Mannella[1,2], Hanno H. Weitering[1,2,*], and Wonhee Ko[1,2,*]*

[1]*Department of Physics and Astronomy, The University of Tennessee, Knoxville, Tennessee 37996, United States*

[2]*Center for Advanced Materials and Manufacturing, The University of Tennessee, Knoxville, Tennessee 37996, United States*

[3]*Department of Materials Science and Engineering, Clemson University, Clemson, South Carolina 29634, United States*

**Abstract.** Monolayer FeSe grown on $SrTiO_3$ (001) exemplifies interface-enhanced superconductivity, exhibiting a transition temperature nearly ten times that of bulk FeSe. Yet, its tunneling spectra lack clearly identifiable superconducting coherence peaks, preventing an unambiguous determination of the gap magnitude. Here, we use ultralow-temperature scanning tunneling spectroscopy as a function of temperature, tip-sample separation, and lateral position to disentangle tunneling contributions by statistical analysis. Decay rates extracted by regression of tip-height-dependent spectroscopy reveal three energy regimes: Andreev-reflection-dominated tunneling below 6 meV, mixed Andreev and quasiparticle tunneling between 6 and 13 meV, and predominantly quasiparticle tunneling above 13 meV. Spatially resolved spectroscopy and correlation analysis further reveal that the tunneling conductance is predominantly localized near defects below 6 meV but becomes spatially extended at higher energies. Together, these results constrain the superconducting gap to lie between 6 and 13 meV. Our findings establish statistical analysis of spectroscopy as a powerful approach to constraining superconducting gap structures when complex spectral features prevent the unambiguous identification of coherence peaks.

Two-dimensional (2D) superconductors provide a versatile platform for investigating how reduced dimensionality, broken symmetries, and interfacial interactions influence superconducting pairing.[1] A particularly striking example is monolayer (ML) FeSe grown on $SrTiO_3$ (STO), which exhibits a superconducting (SC) transition temperature ($T_c$) exceeding 60 K, more than an order of magnitude higher than that of bulk FeSe (~6 K).[2-9] Accordingly, ML FeSe on STO has emerged as a model system for investigating interface-enhanced superconductivity in iron-based materials.[3-10] However, the microscopic origin of this remarkable enhancement remains debated, with interfacial charge transfer and coupling to STO phonons widely considered important contributing factors.[11-

[16] Reported $T_c$ vary with sample preparation and measurement conditions.[9, 17-19] Even the magnitude and structure of the SC gap remain uncertain, with different gap values reported by angle-resolved photoemission spectroscopy (ARPES) and scanning tunneling microscopy/spectroscopy (STM/STS).[9, 10] In particular, the absence of clearly identifiable coherence peaks in tunneling spectra complicates the determination of the SC gap. Resolving these ambiguities is essential for constraining the SC order parameter and ultimately understanding the pairing mechanism.[20-24]

Determining the SC gap structure of ML FeSe is complicated by its multiorbital electronic structure and potentially anisotropic SC order parameter. The gap may depend on both momentum and orbital character, giving rise to different gap magnitudes across the Fermi surface.[24-27] ARPES measurements have reported gap values of approximately 8–13 meV,[26] while STM/STS studies have inferred values ranging from 9 to 17 meV.[20, 22, 28] In STM/STS, two pairs of approximately symmetric peaks near ±9 and ±16 mV have commonly been assigned to SC coherence peaks.[20, 22, 28] However, their unusual spatial dependence and sensitivity to experimental conditions complicate this assignment.[20, 22, 28] Moreover, the spectra lack a well-defined SC density-of-states lineshape that would allow these features to be identified unambiguously. This motivates an alternative approach to constraining the SC gap structure without relying on the assignment of individual spectral peaks to gap-edges.

In this Letter, we investigate the SC gap structure of ML FeSe on STO using ultralow-temperature STM/STS as a function of temperature, tip–sample separation, and lateral position. In particular, the dependence of the tunneling conductance on tip–sample separation allows us to distinguish conventional quasiparticle tunneling from Andreev reflection, providing information that cannot be extracted from the spectral lineshape alone.[29-32] Combined with spatially resolved spectroscopy and correlation analysis, this approach provides complementary constraints on the SC gap structure.[20, 22, 28]

Figure 1 presents representative STM/STS measurements of ML FeSe. The atomically resolved topographic image in Figure 1(a) reveals the square lattice of the surface Se atoms. The high crystalline quality and electron doping nature of the ML FeSe on STO was consistently evidenced by high-energy electron diffraction (RHEED), room-temperature topographic image and ARPES measurement (See Methods, Figures S1, S2). The spatially averaged d*I*/d*V* spectrum acquired over a defect-free region at 4.2 K [green curve in Figure 1(b)] exhibits a U-shaped suppression of conductance with prominent peaks near ±9 and ±16 mV, consistent with previous STM/STS studies.[20, 22, 28] These features have commonly been interpreted as SC coherence peaks. However, at 0.3 K [dark blue curve in Figure 1(b)], additional peaks and shoulders become resolved near ±5, ±6, ±7, and ±13 mV. The resulting complex spectral lineshape raises questions about the assignment of individual peaks to SC gap-edges and complicates a quantitative determination of the gap structure.

To establish that the additional peaks and shoulders are genuine spectral features rather than experimental noise, we measured the temperature dependence of the differential conductance at a fixed location between 0.3 and 10 K [Figure S3(a)]. The spectral evolution is largely reproduced by thermally broadening the 0.3 K spectrum through convolution with the derivative of the Fermi–Dirac distribution [Figure S3(b)]. The measured 0.3 K spectrum was used as the reference and additional thermal broadening were applied to reach each target temperature, accounting for the finite broadening at the base temperature. The effective electron temperature was independently calibrated by fitting the SC gap of Pb (111) using the Dynes model (Figure S4).[33] We also confirmed that the additional spectral features exceed the instrumental noise level determined from measurements on Au (111) (Figure S5). These observations indicate that the fine structure revealed at 0.3 K is not an experimental artifact but becomes less discernible at elevated temperatures primarily because of thermal broadening.

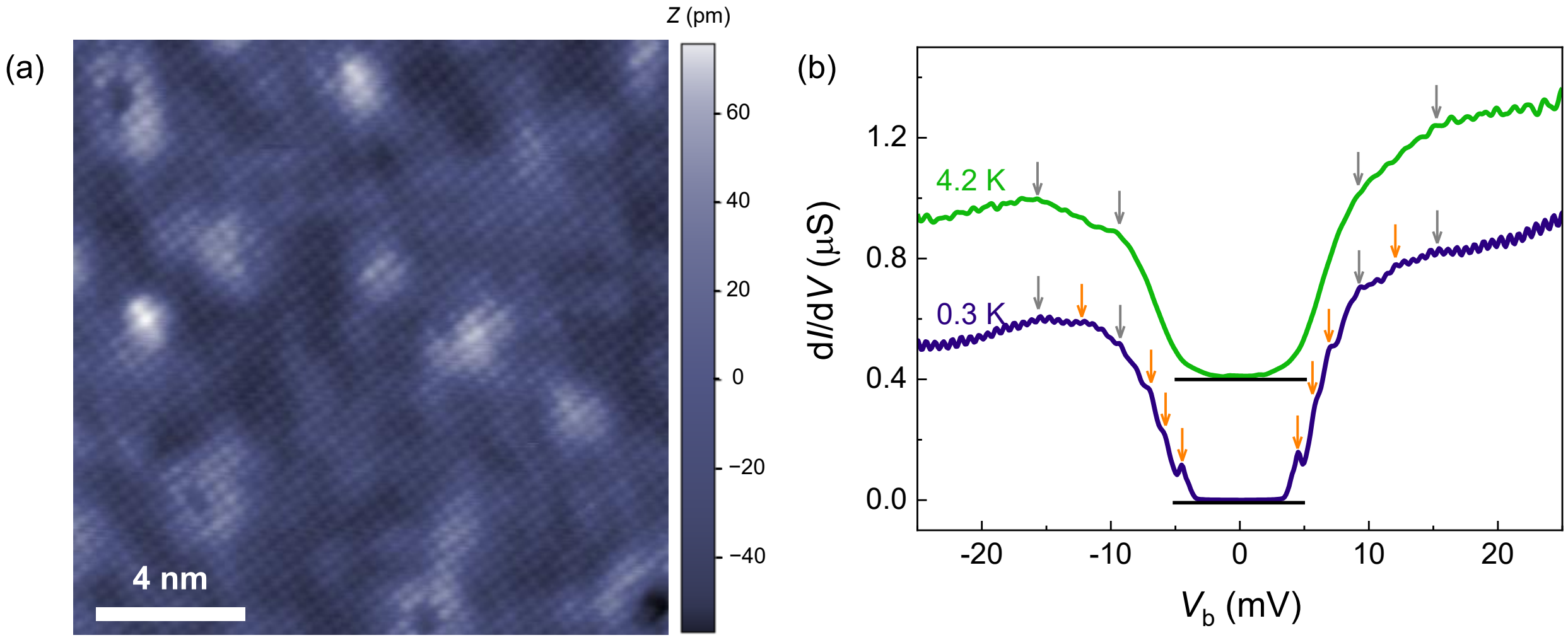

**Figure 1. STM topography and tunneling spectra of ML FeSe on STO.** (a) Atomically resolved STM topography showing the square lattice of surface Se atoms ($V_b$ = 100 mV, $I$ = 50 pA). (b) Spatially averaged d$I$/d$V$ spectra acquired at 0.3 K (dark blue) and 4.2 K (green). The gray arrows indicate prominent spectral peaks near ±9 and ±16 mV, while the orange arrows identify additional peaks and shoulders resolved at 0.3 K. The 4.2 K spectrum is vertically offset for clarity, with the horizontal black lines indicating the respective zero-conductance levels.

To distinguish the contributions of different tunneling processes, we measured the dependence of the differential conductance on tip–sample separation. Figure 2(a) shows the evolution of the spectra at 0.3 K as the tip approaches the surface over a distance of 140 pm. In the tunneling regime, the differential conductance approximately follows an exponential dependence on tip–sample separation,

$$\frac{dI}{dV}(z) \propto e^{-\kappa z}$$

allowing us to extract the decay rate $\kappa$ as a function of bias voltage (Fig. S6).[34, 35] We normalize $\kappa$ to its value at −25 mV, denoted $\kappa_N$, where conventional single-quasiparticle tunneling dominates. As shown in Figure 2(b), the normalized decay rate $\kappa/\kappa_N$ remains close to unity above approximately 13 meV but increases at lower energies, reaching nearly twice its normal-state value below approximately 6 meV. This behavior is consistent with a crossover from conventional single-quasiparticle tunneling to Andreev reflection. In the weak-transmission limit, the Andreev conductance scales approximately as the square of the normal-state tunneling conductance,

$$G_A(z) \propto G_N^2(z) \propto e^{-2\kappa_N z}$$

so that $\kappa_A/\kappa_N \approx 2$, where $\kappa_A$ is decay rate of Andreev reflection.[29, 30, 36, 37] The intermediate values of $\kappa/\kappa_N$ are consistent with contributions from both tunneling processes.[31, 32] Because the measured decay rate is fitted over finite tip-displacement intervals, its value can depend on the selected range when these contributions coexist (Figure S7).

The strongly suppressed differential conductance below approximately 6 meV indicates a negligible density of available quasiparticle states. Therefore, at small tip–sample separations, the conductance exhibits a nearly constant normalized decay rate $\kappa/\kappa_N \approx 2$, consistent with Andreev reflection dominating the tunneling response. Together, these observations are consistent with a fully open SC gap whose minimum magnitude is of order 6 meV, although the measurement is weighted by the tunneling matrix elements. Between 6 and 13 meV (Region II), the normalized decay rate evolves toward the conventional quasiparticle tunneling limit, suggesting contributions from both tunneling processes.[31, 32] Over this energy range, the spectra exhibit a gradual increase in conductance with superimposed peaks and shoulders. These features may reflect the anisotropic

SC density-of-states,[23, 26, 38] coupling to bosonic excitations,[39, 40] or energy exchange with the electromagnetic environment,[41, 42] although their microscopic origin remains unresolved. We therefore interpret the 6–13 meV interval as a SC gap-edge crossover region rather than assigning individual spectral peaks to distinct gap magnitudes. The characteristic energy scales and overall bias dependence of the normalized decay rate are reproduced at different surface locations and over different fitting intervals (Figures S7, S8), supporting the robustness of this interpretation. The quantitative analysis of detailed shape and values of normalized decay rate spectra will be the focus of future work and may reveal the nature of the underlying unconventional superconductivity.[31]

Spatially resolved tunneling spectroscopy provides additional evidence for a characteristic energy scale near 6 meV. Figure 3(a) shows the STM topography of the region in which conductance maps were acquired. At biases below approximately 6 mV, the conductance is strongly suppressed over most of the surface, with the remaining spectral weight predominantly localized near defects [Figure 3(b)]. At higher biases, the conductance becomes more spatially extended, indicating the emergence of delocalized quasiparticle states. This behavior supports the interpretation of the 6 meV thresholds as a possible minimum gap scale: below this energy, quasiparticle tunneling is strongly suppressed except near defects that introduce in-gap states, whereas above it, quasiparticle states become increasingly accessible.[22, 43]

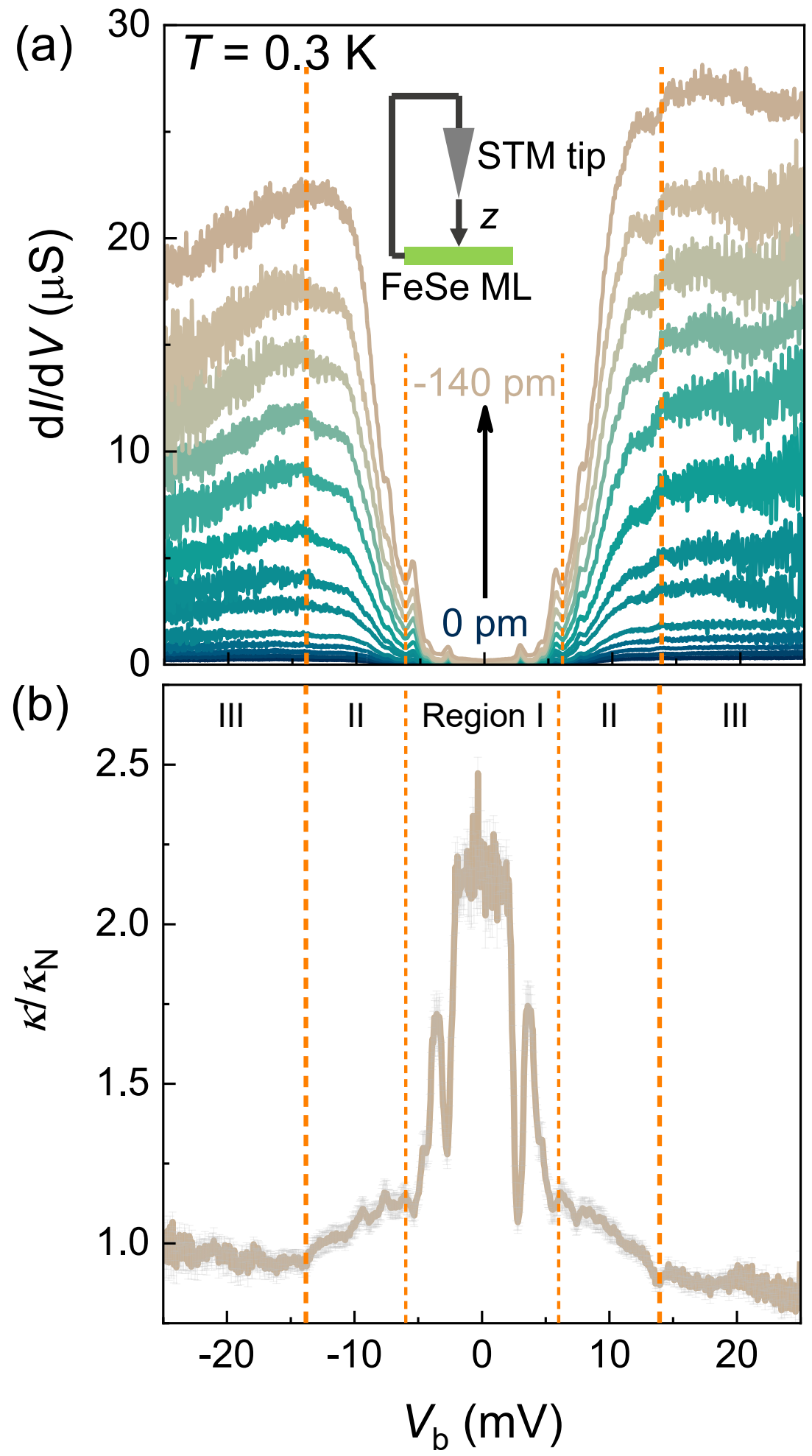

**Figure 2. Tip-height-dependent tunneling spectroscopy of monolayer FeSe on STO.** (a) d$I$/d$V$ spectra acquired at 0.3 K as the tip approaches the sample over a displacement of 140 pm. (b) Normalized decay rate, $\kappa/\kappa_N$, extracted from the exponential tip-height dependence of the conductance, where $\kappa_N$ is the decay constant at −25 mV. Vertical dashed lines mark the characteristic energies near ±6 and ±13 mV, separating three regimes: (I) predominantly Andreev reflection, (II) a crossover with contributions from Andreev reflection and quasiparticle tunneling, and (III) predominantly conventional quasiparticle tunneling.

To further investigate the relationship between spectral features at different energies, we analyze the spatial correlations between conductance maps using the Pearson correlation coefficient. For each pair of bias voltages, $V_{b1}$ and $V_{b2}$, we construct a scatter plot in which each point represents a pixel of the spectroscopic grid, with coordinates given by the differential conductance measured at that position at the two voltages. The Pearson coefficient, $C\ (V_{b1},\ V_{b2})$, is defined as the spatial covariance of the two conductance maps divided by the product of their standard deviations:

$$C(V_{b1}, V_{b2}) = \frac{\langle \delta G(\boldsymbol{r}, V_{b1})\ \delta G(\boldsymbol{r}, V_{b2})\rangle_{\mathrm{r}}}{\sigma(V_{b1})\ \sigma(V_{b2})}$$

Here $G(r, V) = (dI/dV)_{r,V}$, $\delta G(\boldsymbol{r}, V) = G(\boldsymbol{r}, V) - \langle G(\boldsymbol{r}, V)\rangle_{\boldsymbol{r}}$, and $\sigma\ (V)$ denotes the spatial standard deviation, with $\langle ... \rangle_r$ indicating an average over all pixels. The coefficient ranges from −1 for perfect anticorrelation to +1 for perfect positive correlation, with zero indicating no linear correlation. Figure 3(c) shows representative examples of positive, negligible, and negative correlations, while Figure 3(d) presents the correlation coefficient for all pairs of bias voltages. A pronounced negative correlation occurs between conductance below approximately 6 meV and within the 6–13 meV gap-edge region. This anticorrelation indicates that regions with enhanced low-energy conductance, particularly near defects, tend to exhibit suppressed conductance at higher energies, consistent with a redistribution of spectral weight associated with defect-induced in-gap states.[22, 43, 44] In contrast, conductance maps within the gap-edge region exhibit positive spatial correlations, suggesting that their spectral features share a similar spatial dependence. These correlations further support a relationship between defect-induced in-gap states and the spectral features of the SC gap-edge region.

Taken together, the tip-height-dependent and spatially-resolved measurements provide complementary evidence for a SC gap that remains fully open below approximately 6 meV, while revealing a broad gap-edge crossover extending to about 13 meV.[26] Importantly, these characteristic energy scales emerge without assigning any of the prominent spectral peaks to SC coherence peaks.

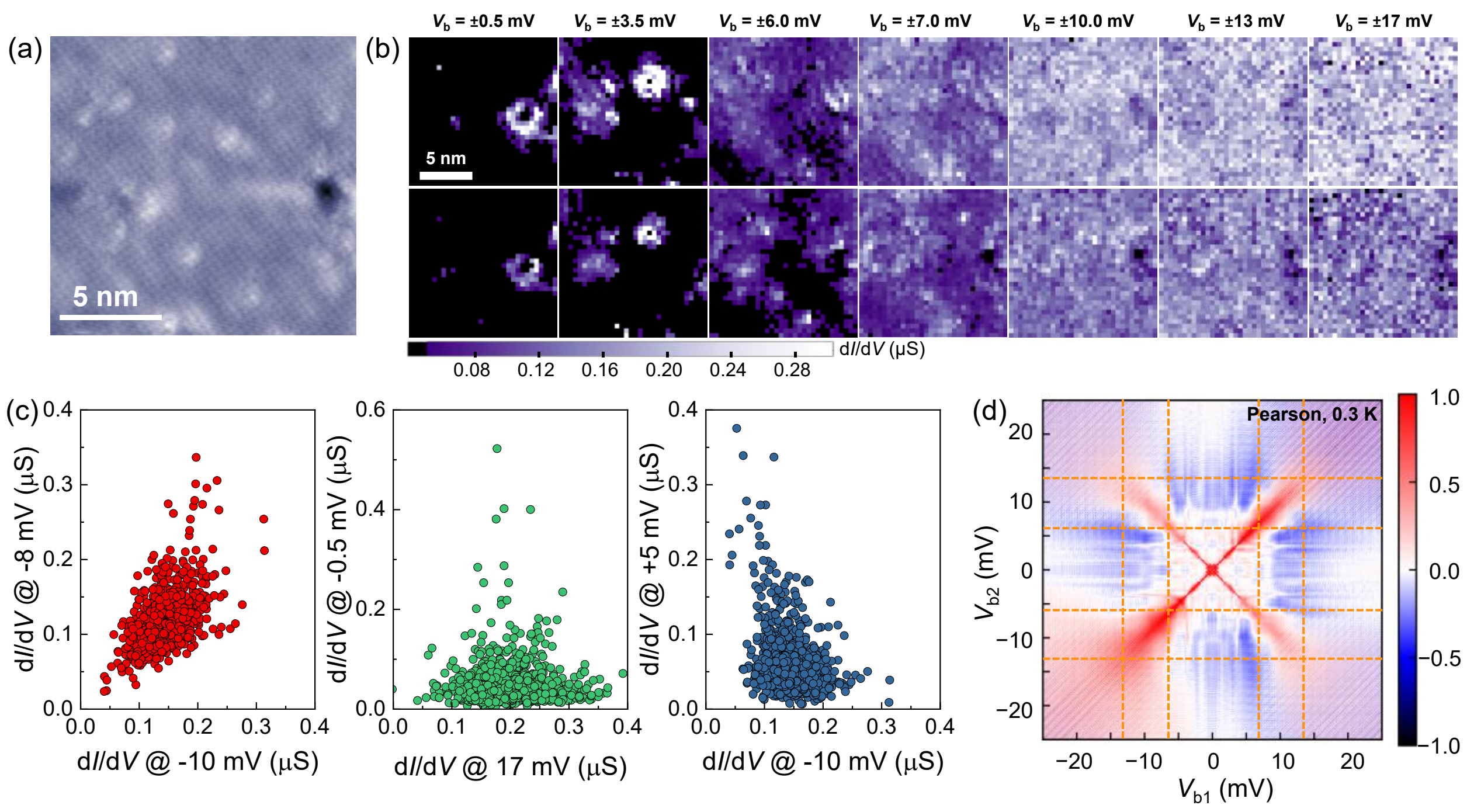

**Figure 3. Spatially resolved tunneling spectroscopy and conductance correlations of ML FeSe on STO.** (a) STM topography of the region investigated ($V_b$ = 25 mV, $I$ = 5 nA). (b) d$I$/d$V$ maps acquired at selected bias voltages between −17 and +17 mV, showing predominantly defect-localized spectral weight at low energies and more spatially extended conductance at higher energies. (c) Representative scatter plots illustrating positive (red), negligible (green), and negative (blue) spatial correlations between conductance maps acquired at different bias voltages. Each point represents one pixel of the spectroscopic grid, with its horizontal and vertical coordinates given by the conductance at the two indicated voltages. (d) Pearson correlation coefficient $C$ ($V_{b1}$, $V_{b2}$) calculated for all pairs of bias voltages $V_{b1}$ and $V_{b2}$. The color scale ranges from −1 (perfect anticorrelation) to +1 (perfect positive correlation). Orange dashed lines indicate the characteristic energy scale near ±6 and 13 mV.

In summary, we have acquired tunneling spectroscopy as a function of temperature, tip-sample separation, and lateral position and analyzed these spectra with several statistical methods to investigate the SC gap structure of ML FeSe on STO. By tracking the evolution from Andreev-reflection-dominated conductance to conventional single-quasiparticle tunneling, we identify distinct energy regimes of 6 and 13 meV that place constraints on the SC gap structure. Combined with spatially resolved spectroscopy, this approach provides a means of characterizing SC gaps when conventional coherence peak assignments are ambiguous. More broadly, tip-height-dependent spectroscopy offers a complementary route to investigating superconductivity in materials with complex tunneling spectra.

**Methods.** ML FeSe was grown by molecular beam epitaxy (MBE) on 0.5 wt% Nb-doped STO (001) substrates. Prior to growth, the substrates were degassed in ultrahigh vacuum (UHV) below 500 °C and annealed at 1000 °C under a Se flux for 45 min to obtain clean, atomically flat surfaces. Substrate quality was verified by RHEED [Figure S1(a)] and room-temperature STM [Figure S1(b)]. FeSe was grown by co-evaporating Fe and Se under Se-rich conditions at a substrate temperature of 450 °C, followed by annealing for 1 h at 480 °C and 17 h at 370 °C. The resulting films were characterized by RHEED [Figure S1(c)] and STM [Figure S1(d)]. ARPES confirmed the characteristic electronic structure of ML FeSe, with electron pockets at the *M* points of the Brillouin zone (Figure S2).[4, 26, 27] For ultralow-temperature STM measurements, the films were capped with amorphous Se below −100 °C to protect them during *ex-situ* transfer to the STM system, and the capping layer was subsequently removed by annealing in UHV. Measurements were performed using a Unisoku USM1300 STM at a base temperature of 0.3 K and a pressure below $5.3 \times 10^{-10}$ mbar. Pt/Ir tips were prepared on Au (111), and d$I$/d$V$ spectra were acquired using lock-in detection with modulation peak amplitudes of 50 μV (Figure 3) and 100 μV (Figures 1, 2) at 873 Hz.

## ASSOCIATED CONTENT

### Supporting Information

The following file is available free of charge.

## AUTHOR INFORMATION

### Corresponding Author

*Hanno H. Weitering (hanno@utk.edu)

*Wonhee Ko (wko@utk.edu)

### Author Contributions

D.S., W.K., and H.W. conceived the research and designed the experiments. D.S. and X.P. synthesized the samples by MBE. K.M. and P.V. performed the ARPES measurements, while D.S. and C.H. carried out ultralow-temperature STM/STS measurements. D.S. analyzed the experimental results. P.M. and W.K. developed the decay-rate analysis. N.M., H.W., and W.K. supervised the project. D.S., H.W., and W.K. wrote the manuscript, which was reviewed by all authors.

**Acknowledgement**

This work was supported by the National Science Foundation through the University of Tennessee MRSEC Center for Advanced Materials and Manufacturing (DMR-2309083). The authors thank Jose Lado, Steven Johnston, and Rui-Xing Zhang for valuable discussions on the theoretical interpretation.

## References

(1) Saito, Y.; Nojima, T.; Iwasa, Y. Highly crystalline 2D superconductors. *Nat. Rev. Mater.* **2016**, *2* (1), 16094.

(2) Hsu, F.-C.; Luo, J.-Y.; Yeh, K.-W.; Chen, T.-K.; Huang, T.-W.; Wu, P. M.; Lee, Y.-C.; Huang, Y.-L.; Chu, Y.-Y.; Yan, D.-C.; et al. Superconductivity in the PbO-type structure *α*-FeSe. *Proc. Natl. Acad. Sci. U. S. A.* **2008** *105* (38), 14262-14264.

(3) Wang, Q.-Y.; Li, Z.; Zhang, W.-H.; Zhang, Z.-C.; Zhang, J.-S.; Li, W.; Ding, H.; Ou, Y.-B.; Deng, P.; Chang, K.; et al. Interface-Induced High-Temperature Superconductivity in Single Unit-Cell FeSe Films on $SrTiO_3$. *Chin. Phys. Lett.* **2012**, *29* (3), 037402.

(4) Liu, D.; Zhang, W.; Mou, D.; He, J.; Ou, Y. B.; Wang, Q. Y.; Li, Z.; Wang, L.; Zhao, L.; He, S.; et al. Electronic origin of high-temperature superconductivity in single-layer FeSe superconductor. *Nat. Commun.* **2012**, *3*, 931.

(5) He, S.; He, J.; Zhang, W.; Zhao, L.; Liu, D.; Liu, X.; Mou, D.; Ou, Y. B.; Wang, Q. Y.; Li, Z.; et al. Phase diagram and electronic indication of high-temperature superconductivity at 65 K in single-layer FeSe films. *Nat. Mater.* **2013**, *12* (7), 605-610.

(6) Tan, S.; Zhang, Y.; Xia, M.; Ye, Z.; Chen, F.; Xie, X.; Peng, R.; Xu, D.; Fan, Q.; Xu, H.; et al. Interface-induced superconductivity and strain-dependent spin density waves in FeSe/$SrTiO_3$ thin films. *Nat. Mater.* **2013**, *12* (7), 634-640.

(7) Peng, R.; Xu, H. C.; Tan, S. Y.; Cao, H. Y.; Xia, M.; Shen, X. P.; Huang, Z. C.; Wen, C. H.; Song, Q.; Zhang, T.; et al. Tuning the band structure and superconductivity in single-layer FeSe by interface engineering. *Nat. Commun.* **2014**, *5*, 5044.

(8) Ge, J. F.; Liu, Z. L.; Liu, C.; Gao, C. L.; Qian, D.; Xue, Q. K.; Liu, Y.; Jia, J. F. Superconductivity above 100 K in single-layer FeSe films on doped $SrTiO_3$. *Nat. Mater.* **2015**, *14* (3), 285-289.

(9) Faeth, B. D.; Yang, S. L.; Kawasaki, J. K.; Nelson, J. N.; Mishra, P.; Parzyck, C. T.; Li, C.; Schlom, D. G.; Shen, K. M. Incoherent Cooper Pairing and Pseudogap Behavior in Single-Layer FeSe/$SrTiO_3$. *Phys. Rev. X* **2021**, *11* (2), 021054.

(10) Huang, D.; Hoffman, J. E. Monolayer FeSe on $SrTiO_3$. *Annu. Rev. Condens. Matter Phys.* **2017**, *8* (1), 311-336.

(11) Zhang, W.; Li, Z.; Li, F.; Zhang, H.; Peng, J.; Tang, C.; Wang, Q.; He, K.; Chen, X.; Wang, L.; et al. Interface charge doping effects on superconductivity of single-unit-cell FeSe films on $SrTiO_3$ substrates. *Phys. Rev. B* **2014**, *89* (6), 060506(R)

(12) Lee, J. J.; Schmitt, F. T.; Moore, R. G.; Johnston, S.; Cui, Y. T.; Li, W.; Yi, M.; Liu, Z. K.; Hashimoto, M.; Zhang, Y.; et al. Interfacial mode coupling as the origin of the enhancement of $T_c$ in FeSe films on $SrTiO_3$. *Nature* **2014**, *515* (7526), 245-248.

(13) Zhang, H.; Zhang, D.; Lu, X.; Liu, C.; Zhou, G.; Ma, X.; Wang, L.; Jiang, P.; Xue, Q. K.; Bao, X. Origin of charge transfer and enhanced electron-phonon coupling in single unit-cell FeSe films on $SrTiO_3$. *Nat. Commun.* **2017**, *8* (1), 214.

(14) Rebec, S. N.; Jia, T.; Zhang, C.; Hashimoto, M.; Lu, D. H.; Moore, R. G.; Shen, Z. X. Coexistence of Replica Bands and Superconductivity in FeSe Monolayer Films. *Phys. Rev. Lett.* **2017**, *118* (6), 067002.

(15) Zhang, C.; Liu, Z.; Chen, Z.; Xie, Y.; He, R.; Tang, S.; He, J.; Li, W.; Jia, T.; Rebec, S. N.; et al. Ubiquitous strong electron-phonon coupling at the interface of FeSe/$SrTiO_3$. *Nat. Commun.* **2017**, *8*, 14468.

(16) Song, Q.; Yu, T. L.; Lou, X.; Xie, B. P.; Xu, H. C.; Wen, C. H. P.; Yao, Q.; Zhang, S. Y.; Zhu, X. T.; Guo, J. D.; et al. Evidence of cooperative effect on the enhanced superconducting transition temperature at the FeSe/$SrTiO_3$ interface. *Nat. Commun.* **2019**, *10* (1), 758.

(17) Zhang, W.-H.; Sun, Y.; Zhang, J.-S.; Li, F.-S.; Guo, M.-H.; Zhao, Y.-F.; Zhang, H.-M.; Peng, J.-P.; Xing, Y.; Wang, H.-C.; et al. Direct Observation of High-Temperature Superconductivity in One-Unit-Cell FeSe Films. *Chin. Phys. Lett.* **2014**, *31* (1), 017401.

(18) Pedersen, A. K.; Ichinokura, S.; Tanaka, T.; Shimizu, R.; Hitosugi, T.; Hirahara, T. Interfacial Superconductivity in FeSe Ultrathin Films on $SrTiO_3$ Probed by In Situ Independently Driven Four-Point-Probe Measurements. *Phys. Rev. Lett.* **2020**, *124* (22), 227002.

(19) Ide, K.; Tanaka, T.; Pedersen, A.; Ichinokura, S.; Hirahara, T. Temperature dependence of the superconducting gap of single-layer FeSe/$SrTiO_3$: Direct comparison between transport and spectroscopic measurements. *Phys. Rev. Mater.* **2022**, *6* (12), 124801

(20) Fan, Q.; Zhang, W. H.; Liu, X.; Yan, Y. J.; Ren, M. Q.; Peng, R.; Xu, H. C.; Xie, B. P.; Hu, J. P.; Zhang, T.; et al. Plain *s*-wave superconductivity in single-layer FeSe on $SrTiO_3$ probed by scanning tunnelling microscopy. *Nat. Phys.* **2015**, *11* (11), 946-952.

(21) Ge, Z.; Yan, C.; Zhang, H.; Agterberg, D.; Weinert, M.; Li, L. Evidence for d-Wave Superconductivity in Single Layer FeSe/$SrTiO_3$ Probed by Quasiparticle Scattering Off Step Edges. *Nano. Lett.* **2019**, *19* (4), 2497-2502.

(22) Zhang, H.; Ge, Z.; Weinert, M.; Li, L. Sign changing pairing in single layer $FeSe/SrTiO_3$ revealed by nonmagnetic impurity bound states. *Commun. Phys.* **2020**, *3* (1), 75

(23) Agterberg, D. F.; Shishidou, T.; O'Halloran, J.; Brydon, P. M. R.; Weinert, M. Resilient Nodeless *d*-Wave Superconductivity in Monolayer FeSe. *Phys. Rev. Lett.* **2017**, *119* (26), 267001.

(24) Hirschfeld, P. J.; Korshunov, M. M.; Mazin, I. I. Gap symmetry and structure of Fe-based superconductors. *Rep. Prog. Phys.* **2011**, *74* (12), 124508.

(25) Peng, R.; Shen, X. P.; Xie, X.; Xu, H. C.; Tan, S. Y.; Xia, M.; Zhang, T.; Cao, H. Y.; Gong, X. G.; Hu, J. P.; et al. Measurement of an enhanced superconducting phase and a pronounced anisotropy of the energy gap of a strained FeSe single layer in $FeSe/Nb:SrTiO_3/KTaO_3$ heterostructures using photoemission spectroscopy. *Phys. Rev. Lett.* **2014**, *112* (10), 107001.

(26) Zhang, Y.; Lee, J. J.; Moore, R. G.; Li, W.; Yi, M.; Hashimoto, M.; Lu, D. H.; Devereaux, T. P.; Lee, D. H.; Shen, Z. X. Superconducting Gap Anisotropy in Monolayer FeSe Thin Film. *Phys. Rev. Lett.* **2016**, *117* (11), 117001.

(27) Zhao, L.; Liang, A.; Yuan, D.; Hu, Y.; Liu, D.; Huang, J.; He, S.; Shen, B.; Xu, Y.; Liu, X.; et al. Common electronic origin of superconductivity in (Li,Fe)OHFeSe bulk superconductor and single-layer $FeSe/SrTiO_3$ films. *Nat. Commun.* **2016**, *7*, 10608.

(28) Tanaka, T.; Akiyama, K.; Yoshino, R.; Hirahara, T. Superconductivity of single unit cell $FeSe/SrTiO_3$ (001): Substrate-surface superstructure dependence. *Phys. Rev. B* **2018**, *98* (12), 121410(R).

(29) Ko, W.; Dumitrescu, E. F.; Maksymovych, P. Statistical detection of Josephson, Andreev, and single quasiparticle currents in scanning tunneling microscopy. *Phys. Rev. Res.* **2021**, *3* (3), 033248.

(30) Ko, W.; Lado, J. L.; Maksymovych, P. Noncontact Andreev Reflection as a Direct Probe of Superconductivity on the Atomic Scale. *Nano. Lett.* **2022**, *22* (10), 4042-4048.

(31) Maksymovych, P.; Song, S. Y.; Lawrie, B.; Ko, W.; Lado, J. L. Fingerprinting Superconductors by Disentangling Andreev and Quasiparticle Currents Across Tunable Tunnel Junctions. *Small* **2026**, *Early View*, e75379.

(32) Ko, W.; Song, S. Y.; Yan, J.; Lado, J. L.; Maksymovych, P. Atomic-Scale Andreev Probe of Unconventional Superconductivity. *Nano. Lett.* **2023**, *23* (17), 8310-8318.

(33) Dynes, R. C.; Narayanamurti, V.; Garno, J. P. Direct Measurement of Quasiparticle-Lifetime Broadening in a Strong-Coupled Superconductor. *Phys. Rev. Lett.* **1978**, *41* (21), 1509-1512.

(34) Tersoff, J.; Hamann, D. R. Theory and Application for the Scanning Tunneling Microscope. *Phys. Rev. Lett.* **1983**, *50* (25), 1998-2001.

(35) Tersoff, J.; Hamann, D. R. Theory of the scanning tunneling microscope. *Phys Rev B* **1985**, *31* (2), 805-813.

(36) Blonder, G. E.; Tinkham, M.; Klapwijk, T. M. Transition from metallic to tunneling regimes in superconducting microconstrictions: Excess current, charge imbalance, and supercurrent conversion. *Phys. Rev. B* **1982**, *25* (7), 4515-4532.

(37) Sukhachov, P. O.; von Oppen, F.; Glazman, L. I. Andreev Reflection in Scanning Tunneling Spectroscopy of Unconventional Superconductors. *Phys. Rev. Lett.* **2023**, *130* (21), 216002.

(38) Eugenio, P. M.; Vafek, O. Classification of symmetry derived pairing at the *M* point in FeSe. *Phys. Rev. B* **2018**, *98* (1), 014503.

(39) Liu, C.; Wang, Z.; Ye, S.; Chen, C.; Liu, Y.; Wang, Q.; Wang, Q. H.; Wang, J. Detection of Bosonic Mode as a Signature of Magnetic Excitation in One-Unit-Cell FeSe on $SrTiO_3$. *Nano. Lett.* **2019**, *19* (6), 3464-3472.

(40) Butler, C. J.; Yoshida, M.; Hanaguri, T.; Iwasa, Y. Doublonlike Excitations and Their Phononic Coupling in a Mott Charge-Density-Wave System. *Phys. Rev. X* **2021**, *11* (1), 011059.

(41) Peters, O.; Bogdanoff, N.; Acero González, S.; Melischek, L.; Simon, J. R.; Reecht, G.; Winkelmann, C. B.; von Oppen, F.; Franke, K. J. Resonant Andreev reflections probed by photon-assisted tunnelling at the atomic scale. *Nat. Phys.* **2020**, *16* (12), 1222-1226.

(42) Ast, C. R.; Jack, B.; Senkpiel, J.; Eltschka, M.; Etzkorn, M.; Ankerhold, J.; Kern, K. Sensing the quantum limit in scanning tunnelling spectroscopy. *Nat. Commun.* **2016**, *7*, 13009.

(43) Balatsky, A. V.; Vekhter, I.; Zhu, J.-X. Impurity-induced states in conventional and unconventional superconductors. *Rev. Mod. Phys.* **2006**, *78* (2), 373-433.

(44) Ohnmacht, D. C.; Belzig, W.; Cuevas, J. C. Full counting statistics of Yu-Shiba-Rusinov bound states. *Phys. Rev. Res.* **2023**, *5* (3), 033176.

## Table of Contents

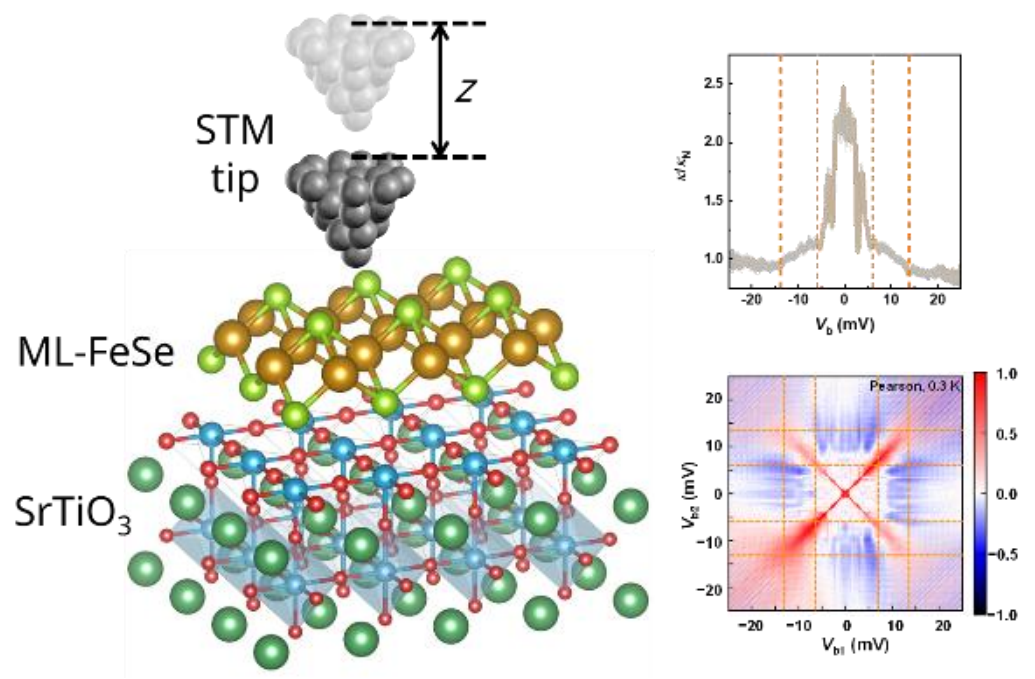